\documentclass[a4paper,11pt]{article}
\usepackage{pos}

\usepackage{ifpdf}

\newcommand{\Nf}{N_\text{f}}
\newcommand{\Nt}{N_\tau}
\newcommand{\Z}{Z_2}
\newcommand{\clqcd}{CL\kern-.25em\textsuperscript{2}QCD \;}
\newcommand{\bb}{\ensuremath{\beta}}

\pdfoutput=1

\title{The chiral phase transition from strong to weak coupling}
\author[a]{Francesca Cuteri}
\author*[a,b]{Alfredo D'Ambrosio}
\author[a,b]{Owe Philipsen}
\author[a]{Alessandro Sciarra}

\affiliation[a]{Institut für Theoretische Physik - Goethe-Universität Frankfurt am Main,\\
  Max-von-Laue-Str. 1, 60438 Frankfurt am Main, Germany}

\affiliation[b]{John von Neumann Institute for Computing (NIC) GSI, \\
 Planckstr. 1, 64291 Darmstadt, Germany}

\emailAdd{ambrosio@itp.uni-frankfurt.de}
\emailAdd{cuteri@itp.uni-frankfurt.de}
\emailAdd{philipsen@itp.uni-frankfurt.de}
\emailAdd{sciarra@itp.uni-frankfurt.de}

\abstract{The order of the chiral phase transition of lattice QCD with unimproved 
staggered fermions is known to depend on the number of quark flavours, their masses and the 
lattice spacing. Previous studies in the literature for $\Nf \in \{ 3,4 \}$ show first-order transitions, 
which weaken with decreasing lattice spacing. Here we
investigate what happens when lattices are made coarser to establish contact to the strong coupling region.
For $\Nf \in \{4,8 \}$ we find a drastic weakening of the transition when going from $\Nt=4$ to $\Nt=2$, which is 
consistent with a second-order chiral transition reported in the literature for $\Nf=4$ in 
the strong coupling limit. This implies a non-monotonic behaviour of the critical quark or
pseudo-scalar meson mass, which separates first-order transitions from crossover behaviour, as a function
of lattice spacing.}

\FullConference{%
 The 38th International Symposium on Lattice Field Theory, LATTICE2021
  26th-30th July, 2021
  Zoom/Gather@Massachusetts Institute of Technology
}

\begin{document}
\maketitle

\begin{figure}[t]
\begin{minipage}[t]{0.4\linewidth}
    \centering
    \includegraphics[width=1\textwidth]{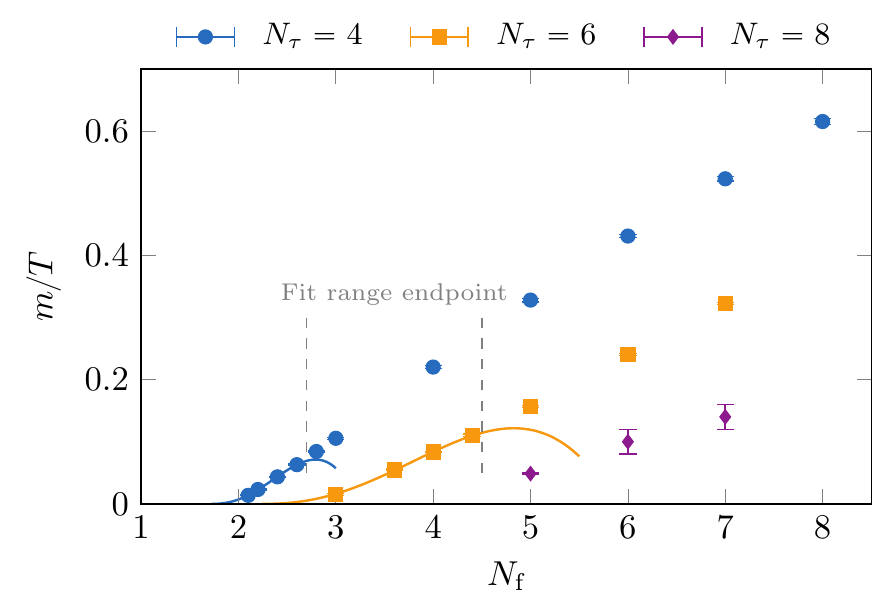}
    \caption{Critical $\Z$ mass  as a function of the number of degenerate staggered quark flavours for different lattice spacings at zero density, in the weak coupling regime \cite{3}.}
    \label{weak}
\end{minipage}
\hspace{1.3cm}
\begin{minipage}[t]{0.5\linewidth}
    \centering
    \includegraphics[width=1\textwidth]{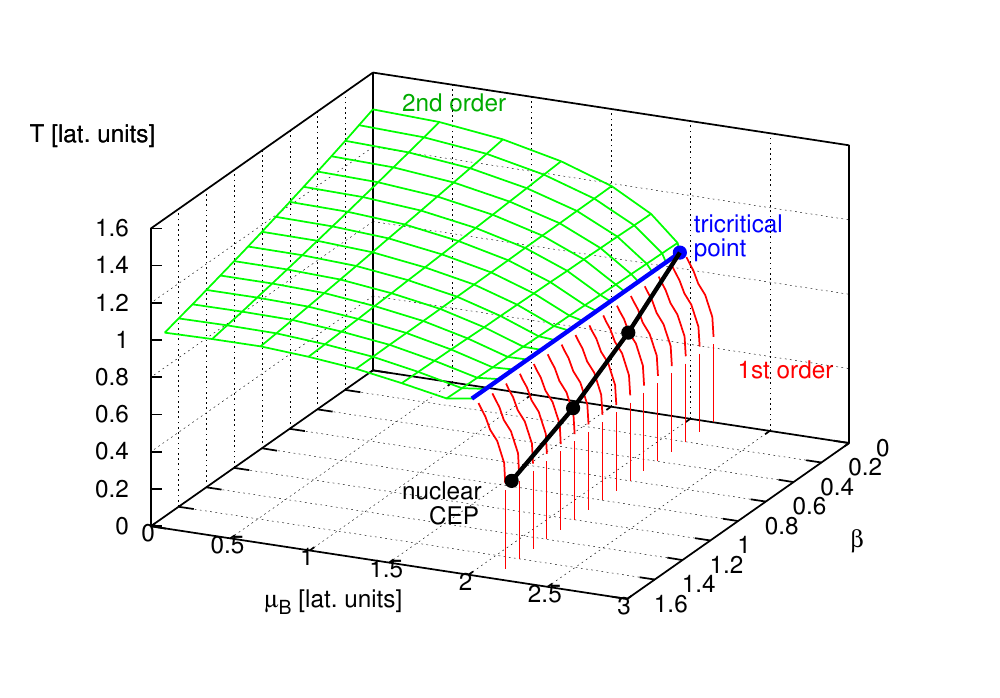}
    \caption{Order of the chiral phase transition using $\Nf=4$ at zero density in strong coupling regime, using staggered fermions \cite{8}.}
    \label{strong}
\end{minipage}
\end{figure}

\section{Introduction}
The order of the chiral phase transition in the chiral limit is still a challenging topic of lattice QCD. Simulations at zero mass on the lattice are prohibited by the influence of zero modes of the Dirac operator, and extrapolations to the chiral limit are necessary in order to draw any conclusions.
In the weak coupling regime, the chiral phase transition at zero chemical potential has been investigated for different numbers of quark flavours and masses, using unimproved staggered fermions and lattice temporal extents $N_\tau$ \cite{1,2,3}. In figure \ref{weak}, the second-order critical boundary line separating 
the parameter region with first-order transitions (below) from crossover (above) is 
shown for $\Nt \in \{ 4,6 \}$. For such studies, we consider a statistical system at zero density described by 
the partition function
\begin{equation}                                                                                                                                                                                                                                                                                                                                                                                                                                                                                                                                                                                                                                                                                                                                                                                                                                                                                                                                     
Z(\beta,am,\Nf,\Nt) = \int \mathcal{D} U (\det D[U])^{\Nf} e^{-S_g[U]},                                                                                                                                                                                                                                                                                                                                                                                                                                                                                                                                                                                                                                                                                                                                                                                                                                                                   
\label{eqn:2}
\end{equation}
in which $\Nf$ can be generalised to continuous values. This is implemented 
straightforwardly for rooted staggered fermions.
For sufficiently small masses, $\Nf$ can then be used as extrapolation parameter with known scaling behaviour, since it features a tricritical point in the lattice chiral limit, where the chiral phase transition changes from first order triple to second order \cite{1}. The difference between the data at different $\Nt$ values shows a strong cutoff dependence, 
which increaes with $\Nf$. The values of the gauge coupling parameter along the critical lines are in the range $\bb\in[4.7, 5.2]$, which means that the data belong to the intermediate to weak coupling regime.

On the other hand, the strong coupling limit for $\Nf=4$ has been thoroughly studied over the last forty years, using both the mean field approximation \cite{4,5,6,7} and Monte Carlo simulations \cite{8,9,10,11}. The latter employ
a reformulation of the partition function in terms of monomers and polymers \cite{9,10}, which permits
a direct simulation of the massless limit. 
Furthermore, the leading $\mathcal{O}(\beta)$
gauge corrections have been included in \cite{12}. 
In figure \ref{strong} we see that, moving towards the strong coupling regime, 
the phase transition at $\mu=0$ is found to be of second order. Since the lattice temporal extent 
$\Nt$ is related to $\bb$ through the lattice spacing $a$,
\begin{equation}                                                                                                                                                                                                                                                                                                                                                                                                                                                                                                                                                                                                                                                                                                                                                                                                                                                                                                                                    
\Nt = ( a(\bb) T)^{-1},  \quad \bb= 6/(g(a))^2,                                                                                                                                                                                                                                                                                                                                                                                                                                                                                                                                                                                                                                                                                                                                                                                                                                                                                          
\end{equation}
\label{eqn:1}a smaller $\Nt$ corresponds to a larger $a$ and smaller $\bb$ at fixed temperature $T$. Therefore, comparing figure \ref{strong} to results for $\Nf = 4$ in figure \ref{weak}, 
we expect the size of the first-order region to shrink again when using a coarser lattice, i.e.~when moving 
towards stronger couplings.

In this work we are trying to establish  the shape of the $\Z$ boundary line between the weak and the strong coupling regimes for $\Nf \in \{ 4,8 \}$. To this end, we perform simulations on $\Nt=2$ lattices 
at different bare quark masses to identify the critical $am_{\Z}$ mass. 

\section{Analysis}

\begin{figure}[t]
\begin{minipage}[t]{0.5\linewidth}
    \centering
    \includegraphics[width=1\textwidth]{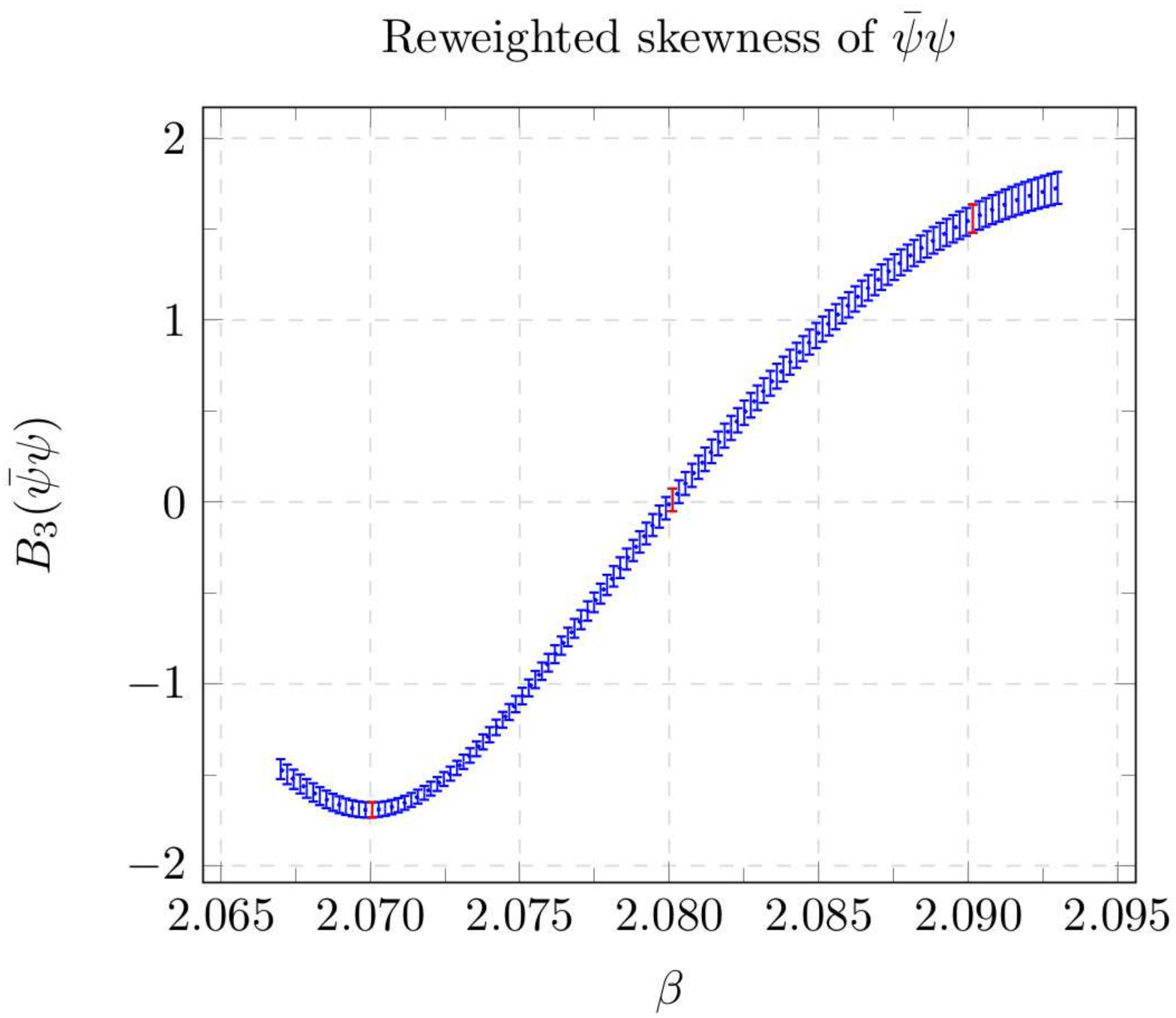}
\end{minipage}
\hspace{0.3cm}
\begin{minipage}[t]{0.5\linewidth}
    \centering
    \includegraphics[width=1\textwidth]{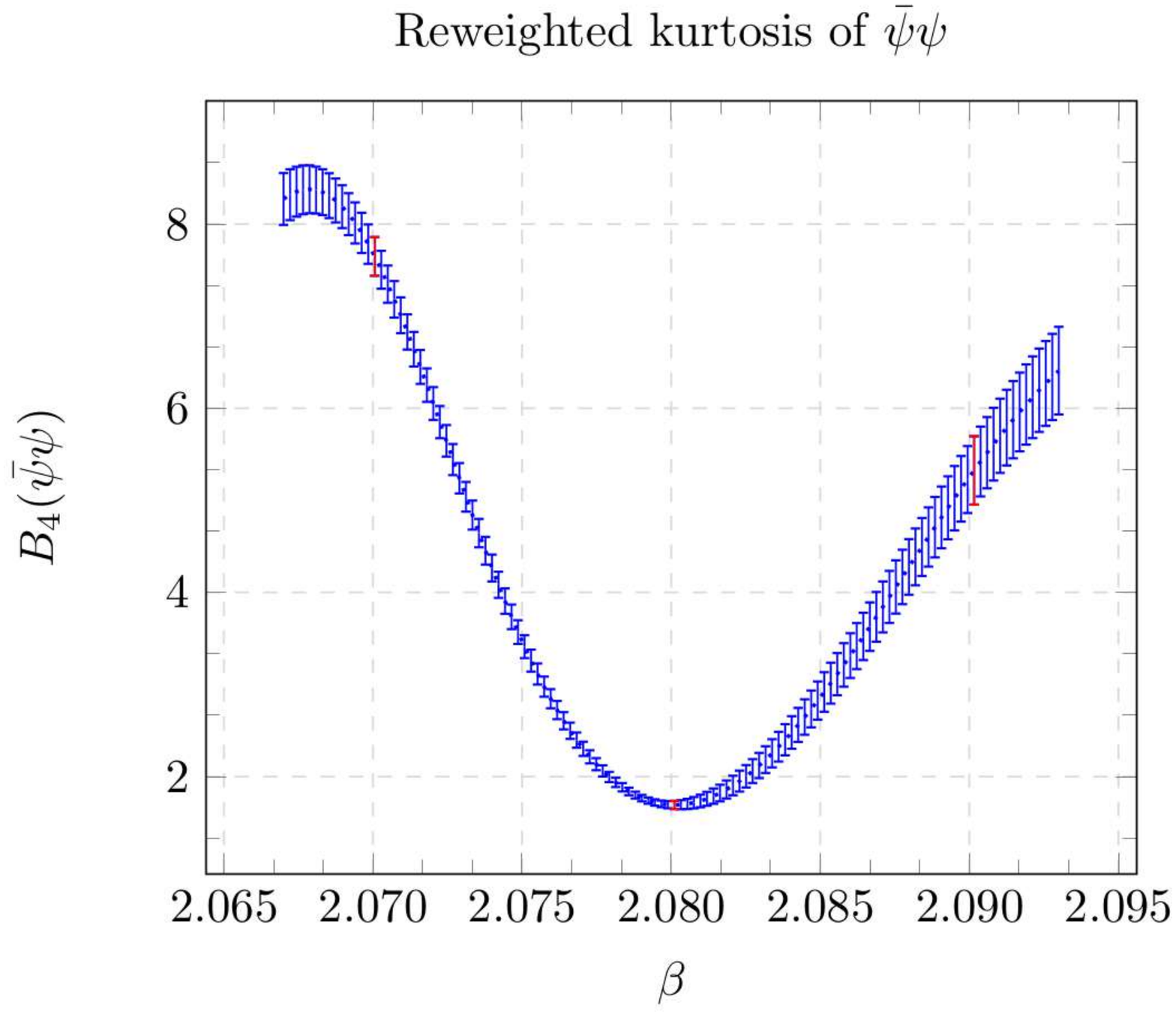}
\end{minipage}
\caption{Example of reweighted skewness and kurtosis for $\langle \bar{\psi} \psi \rangle$ as a function of the gauge coupling $\beta$ for $\Nf=8$, $am = 0.0015$ and $N_\sigma=16$. The red points correspond to the simulated $\beta$ values while the rest has been produced using the Ferrenberg-Swendsen reweighting. }
\label{Skew}
\end{figure}

In order to locate the phase transition and identify its nature, we use the chiral condensate $\langle \bar{\psi} \psi \rangle$, which becomes  a true order parameter in the limit of massless quarks, whereas
it is always non-zero for non-vanishing quark masses. Nevertheless, the sampled distribution of 
$\langle \bar{\psi} \psi \rangle$ contains all information to characterise the phase transition.
We analyse the standardised moments of $\mathcal{O}=\langle \bar{\psi} \psi \rangle$, which for 
fixed $\Nf$ and $\Nt$ read
\begin{equation}                                                                                                                                                                                                                                                                                                                                                                                                                                                                                                                                                                                                                                                                                                                                                                                                                                                                                                                                     
 B_n(\bb,am) = \frac{\left\langle\left(\mathcal{O} - \left\langle\mathcal{O}\right\rangle\right)^n\right\rangle}{\left\langle\left(\mathcal{O} - \left\langle\mathcal{O}\right\rangle\right)^2\right\rangle^{n/2}}                                                                                                                                                                                                                                                                                                                                                                                                                                                                                                                                                                                                                                                                                                                                    .
\label{eqn:4}
\end{equation}

\begin{table}
\centering
\begin{tabular}{ |c|c|c|c| } 
 \hline
  & Crossover & $1^{st}$ order & $3D$ Ising\\ 
 \hline
 $B_4$ & $3$ & $1$ & $1.604$ \\ 
 $\nu$ & $-$ & $1/3$ & 0.6301(4)\\ 
 \hline
\end{tabular}
\caption{Universal infinite volume values of the kurtosis $B_4$ and of the relevant critical exponent.}
\label{tab:Table}
\end{table}
The (pseudo-)critical coupling $\beta_c$, corresponding to the location of the phase boundary, is defined by $B_3(\beta_c,am)=0$, where $B_3(\beta,am)$ is the skewness, which quantifies the asymmetry of the distribution of the order parameter. The order of the phase transition can be investigated by means of the kurtosis $B_4(\beta, am)$, which has
a minimum at the phase boundary $\beta_c$.  Its infinite volume values along the phase boundary, $B_4(\beta_c, am)$, distinguish the different cases and are listed in table \ref{tab:Table}.  When finite volumes are simulated, the kurtosis varies smoothly between the tabulated values as the quark mass is changed. For large enough volumes $N_\sigma^3$, and in a neighbourhood of a critical point, 
the kurtosis evaluated  at $\beta_c$ can be expanded in the scaling 
variable $(am - am_{\Z}) N_\sigma^{1/\nu}$, with a critical exponent given in table \ref{tab:Table},
\begin{equation}
B_4(\beta_c, am, N_\sigma) \approx B_4 (\beta_c, am, \infty) + c (am -am_{\Z})N_\sigma^{1/\nu}.
\label{eqn:scaling}
\end{equation}
The critical mass is therefore defined as the intersection point of the kurtosis evaluated on different volumes.

In principle $B_3$ and $B_4$ are continuous functions of the gauge coupling parameter,
but simulations are carried out for two or three \bb-values per mass of interest. The Ferrenberg-Swendsen reweighting is then employed to improve the resolution in the determination of $\beta_c$, 
as shown in figure \ref{Skew}. 
In our simulations, we fix the lattice temporal extent to the coarsest possible choice $\Nt = 2$. 
All simulations are performed by means of the RHMC algorithm using different numbers of pseudofermions \cite{13} , which allows the algorithm to be used also if $\Nf( \mathrm{mod}4)=0$, and it is implemented in the publicly available simulation program \clqcd based on OpenCL, in its latest version 1.1 \cite{14,15}. For handling effectively the needed simulations the BaHaMAS software \cite{16} is employed.

\section{Results}

 \begin{figure}[t]
\center
\includegraphics[width=1\linewidth,clip]{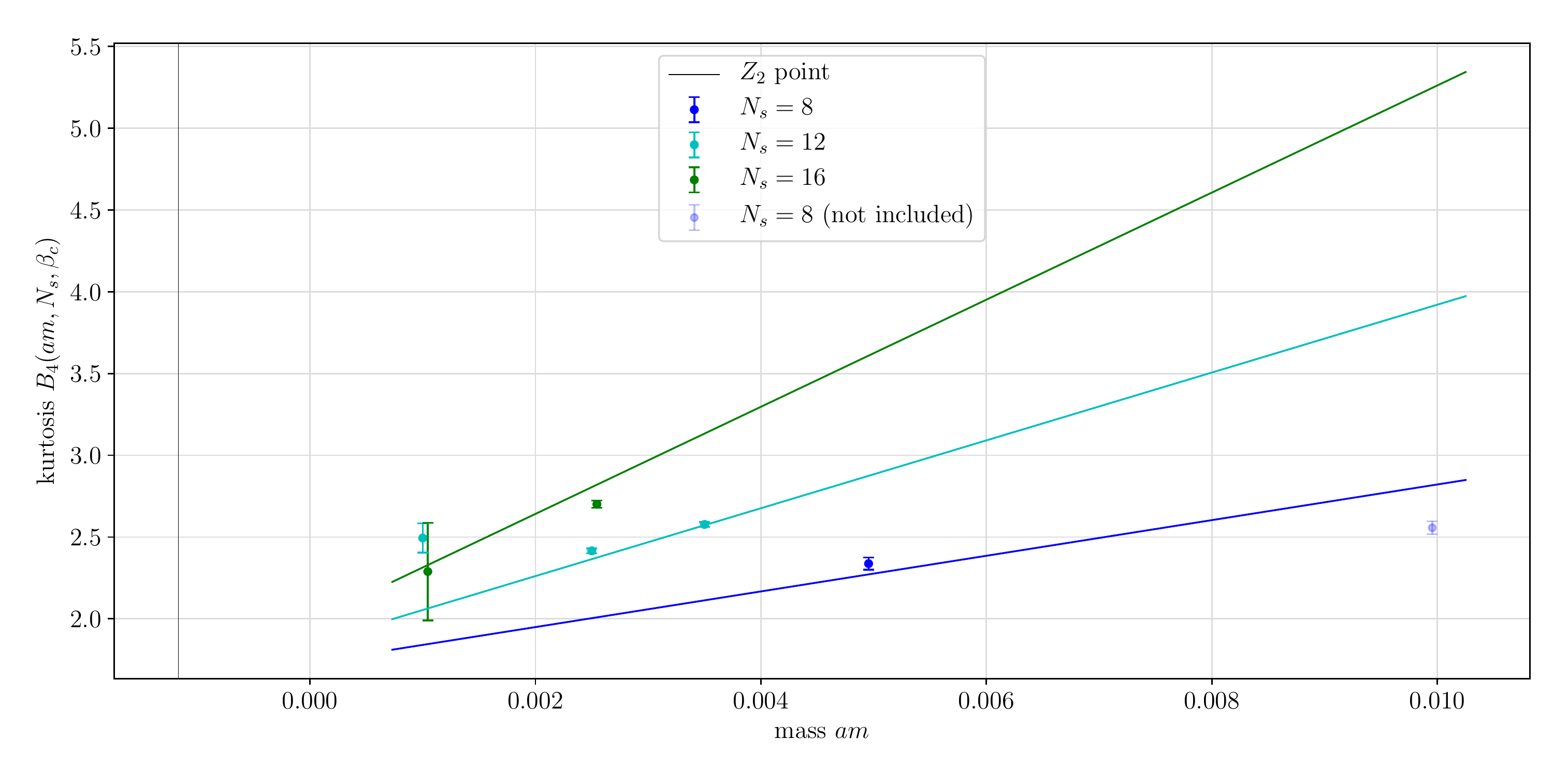}
\caption{Fit of minimum of the kurtosis as a function of the quark masses for $\Nf = 4$.}  
\label{nf4}
\end{figure}
First preliminary results for $\Nf = 4$ are shown in figure \ref{nf4}, using $N_\sigma \in \{ 8, 12, 16 \}$ and simulating 
four different mass values. All data points in figure \ref{nf4}  have $B_4>2$, and 
moreover, according to the fit attempt to equation (\ref{eqn:scaling}), $B_4$ grows with increasing volume. Comparing with
table \ref{tab:Table} we see that those values are far from a $3D$ Ising critical point, and on the crossover side of it.
If we nevertheless attempt linear fits according to equation (\ref{eqn:scaling}),  
no crossing occurs for $am>0$. Thus the critical mass for $\Nf=4$ on $\Nt=2$ is too small to be detected in our simulations,
\begin{equation}
am_{\Z} < 0.001\;.
\end{equation}
Unfortunately, simulations with $am<0.001$ are prohibitively difficult. The sampled chiral condensate distribution features finite size effects when simulating too small masses for a fixed volume. Consequently, the shape of the skewness as a function of $\beta$ undergoes distortions, affecting the scaling analysis. A detailed overview of this problem has been presented in section IV of \cite{1}. On the other hand, larger volumes increase the simulation time beyond the capacities of our current resources. 

Nevertheless, if the absence of a critical $am_{\Z}$ persists once more simulation points and statistics are added, 
it is consistent with
the transition in the lattice chiral limit being of second order. 
The gauge coupling values along the phase boundary are $\beta \geq 3.45$, which is still far from the strong 
coupling limit. This indicates that the second-order transition at $\mu=0$ in figure \ref{strong} extends to these
intermediate coupling values.

\begin{figure}[t]
\center
\includegraphics[width=1\linewidth,clip]{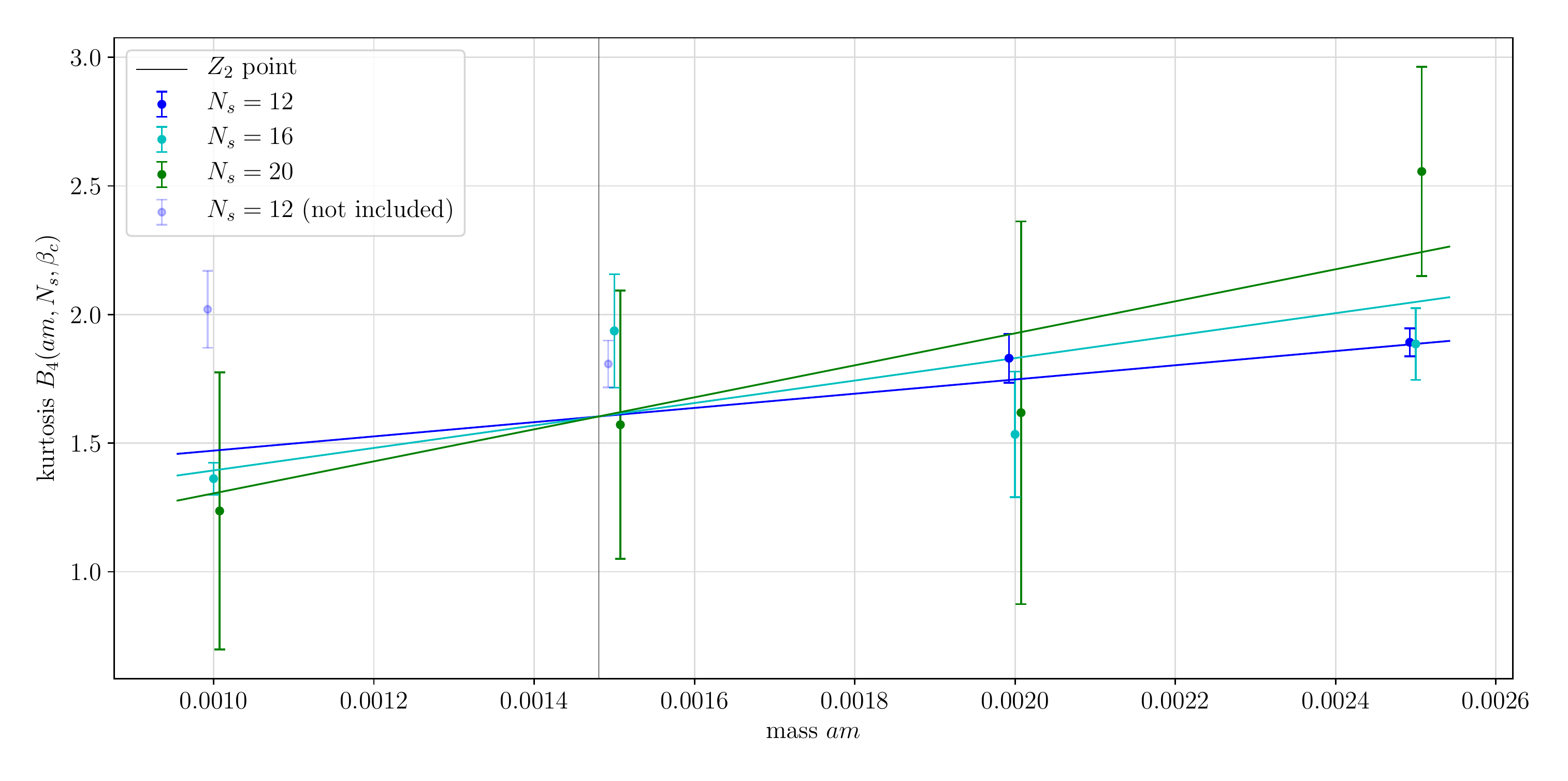}
\caption{Fit of minimum of the kurtosis as a function of the quark masses for $\Nf = 8$.}  
\label{nf8}
\end{figure}

The idea then is to move to higher values of $\Nf$. Figure \ref{weak} shows that the transition generally strengthens
with $\Nf$, so we expect to find a critical point for a certain value of $\Nf > 4 $. In order to increase the chances to succeed, we performed new simulations for $\Nf=8$, with results as shown in figure \ref{nf8}.
We  have used larger volumes $N_\sigma \in \{ 12,16,20 \}$ in order to minimise the influence from finite size effects for small masses. Nevertheless, the data points corresponding to $am \in \{ 0.001, 0.0015 \}$ and $N_\sigma=12$, although presented in the plot, are not included in the fit since their larger values are again strongly affected by finite size effects \cite{1}.

As expected, an  important difference to the fit in figure \ref{nf4} is that we are now able to locate a critical mass 
\begin{equation*}
am_{\Z}=0.00148(10),
\end{equation*}
where the three lines in the fit do cross. The values of  $\beta_c\in[2.06, 2.105]$ are
closer to the strong coupling regime than in the $\Nf=4$ case. 
Since some of the data were not included in the fit, we are performing additional simulations for $am \in \{0.001, 0.0015 \}$  using the larger volume $N_\sigma=24$, but so far the statistics is not enough for a reliable analysis. Low statistics also explains the larger error bars for the larger volumes in figure \ref{nf8}, as well as for the large uncertainty on the first $am_{\Z}$ estimate.

\section{Conclusions}
\begin{figure}[t]
\centering
    \includegraphics[width=0.49\textwidth]{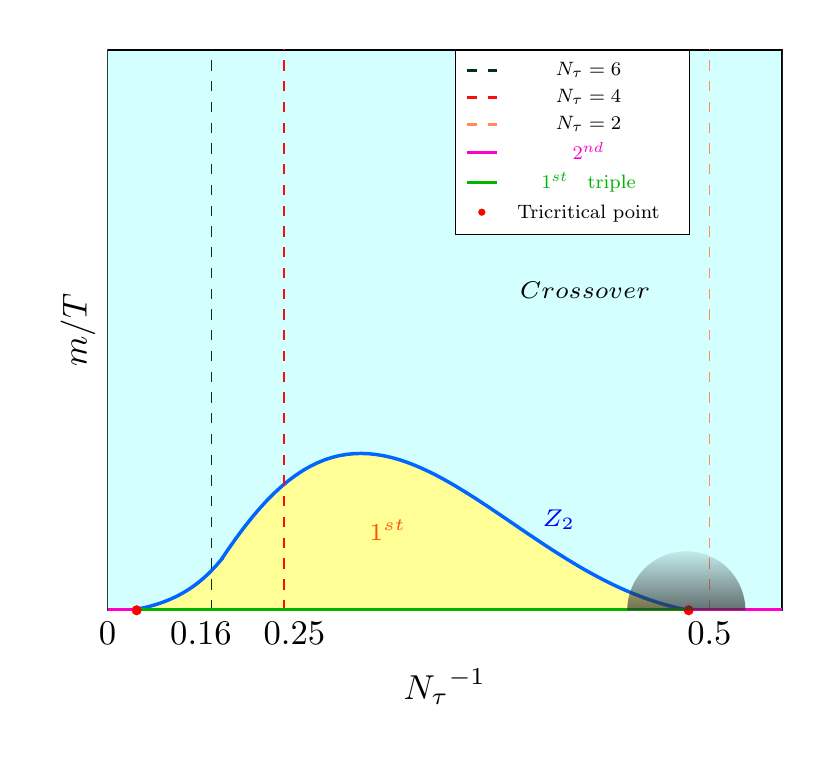}
    \includegraphics[width=0.49\textwidth]{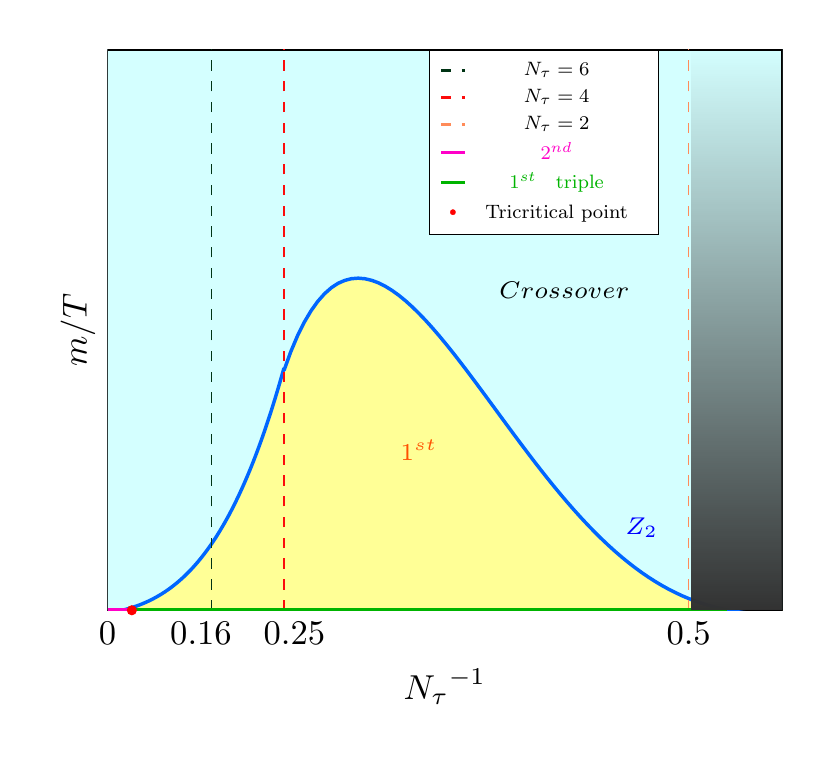}
\caption{Sketch of the $(m/T,\Nt^{-1})$-Columbia Plot for $\Nf=4$ (left) and $\Nf=8$ (right). Red dots represent tricritical points, where the first-order triple $am=0$  transition line meets the second-order ones. The shaded grey areas remain unknown. The rightmost dashed line represents $\Nt=2$ simulated in this work, the other two correspond to $\Nt \in \{ 4,6 \}$ from \cite{2,3}. The $\Nt=6$ line for $\Nf=8$ does not represent a numerical result but is kept there to give an idea of how much the first-order region differs from the one of $\Nf=4$.} 
\label{sketch}
\end{figure}

In figure \ref{sketch} we present two qualitative sketches of the critical $\Z$ line for $\Nf=4$ and $\Nf=8$ based
on our preliminary results. According to \cite{3}, both $\Nf \in \{4,8\} $ theories most likely display a tricritical point
$\Nt^\mathrm{tric}(\Nf)$ on the horizontal axis for large $\Nt$, which is closer to the origin, the larger $\Nf$.
The trend of the boundary lines between $\Nt=6$ and $\Nt=4$ is based on figure \ref{weak} and also in agreement with \cite{17}. Our preliminary data show that towards larger $\Nt^{-1}$, i.e.~stronger couplings, the critical lines 
for both cases is coming down again, and for $\Nf=4$ it must terminate in another tricritical point, which is
consistent with $4<\Nt^\mathrm{tric}<2$. Since no $am_{\Z}$ was found, the location of the tricritical point remains open. This is indicated by the shaded area in the sketch.
On the other hand, moving  to $\Nf=8$ it was possible to locate a candidate critical mass, represented by the intersection of the rightmost dashed line and the boundary line in figure  \ref{sketch}. In this case, the fate of the transition in the strong coupling limit is not known.

Altogether, we find that the order of the chiral phase transition
changes non-monotonically as a function of lattice spacing for both $\Nf \in \{4,8 \} $ unimproved staggered quarks, 
which highlights the requirement of the correct ordering of taking 1) the continuum limit and 2)  the chiral limit
before any conclusions for continuum physics can be drawn. 
 \\ 

\acknowledgments
All simulations have been performed on the Goethe-HLR cluster at Goethe University Frankfurt. The authors acknowledge support by the Deutsche Forschungsgemeinschaft (DFG, German Research Foundation) through the CRC-TR 211 'Strong-interaction matter under extreme conditions'– project number 315477589 – TRR 211, and
by the Helmholtz Graduate School for Hadron and Ion Research (HGS-HIRe) .


\begin{thebibliography}{99}
\bibitem{1}
F. Cuteri, O. Philipsen, and A. Sciarra,
{\emph{The QCD chiral phase transition from non-integer numbers of flavors}},
\href{https://doi.org/10.1103/PhysRevD.97.114511} {\emph{Phys. Rev. D \textbf{97} (2018)}}
[{\tt arxiv:1711.05658[hep-lat]}]


\bibitem{2}
F. Cuteri, O. Philipsen and A. Sciarra,
\emph{Progress on the nature of the QCD thermaltransition as a function of quark flavors and masses},
\href{https://doi.org/10.22323/1.334.0170} {\emph{PoS \textbf{LATTICE2018} (2018)}}
[{\tt arxiv:1811.03840[hep-lat]}]

\bibitem{3}
F. Cuteri, O. Philipsen, A. Sciarra, 
\emph{On the order of the QCD chiral phase transition for different numbers of quark flavours} ,
\href{https://doi.org/10.1007/JHEP11(2021)141}{\emph{J. High Energ. Phys.  } (2021) 141}
[{\tt   	arXiv:2107.12739 }]

\bibitem{4}
N. Bilić, F. Karsch, K. Redlich,
    \emph{ Flavor dependence of the chiral phase transition in strong-coupling QCD},
    \href{https://doi.org/10.1103/PhysRevD.45.3228}{\emph{Phys. Rev. D} (1992) 45}

\bibitem{5}
N. Kawamoto, J. Smit,
    \emph{ Effective lagrangian and dynamical symmetry breaking in strongly coupled lattice QCD},
    \href{https://doi.org/10.1016/0550-3213(81)90196-6}{\emph{Nuclear Physics B} (1981) 192}

\bibitem{6}
P.H. Damgaard, N. Kawamoto, K. Shigemoto,
    \emph{ Strong Coupling Analysis of the Chiral Phase Transition at Finite Temperature},
    \href{https://doi.org/10.1016/0550-3213(86)90470-0}{\emph{Nuclear Physics B} (1986) 264}

\bibitem{7}
F. Karsch, H. W. Wyld
    \emph{ Complex Langevin Simulation of the SU(3) Spin Model with Nonzero Chemical Potential},
    \href{https://doi.org/10.1103/PhysRevLett.55.2242}{\emph{Phys. Rev. Lett.}, (1985) 55} 

\bibitem{8}
P. de Forcrand, O. Philipsen and W. Unger,
\emph{QCD phase diagram from the lattice at strong coupling},
\href{https://doi.org/10.22323/1.217.0073} {\emph{PoS \textbf{CPOD2014} (2015)}}
[{\tt arxiv:1503.08140[hep-lat]}]

\bibitem{9}
Karsch, F. and Mutter, K. H.,
    \emph{Strong coupling QCD at finite baryon number density},
    \href{https://doi.org/10.1016/0550-3213(89)90396-9}{\emph{Nuclear Physics B} \textbf{313} 541--559 (1989)}
    
\bibitem{10}
Rossi, P. and Wolff, U,
    \emph{ Lattice QCD with fermions at strong coupling: A dimer system},
    \href{https://doi.org/10.1016/0550-3213(84)90589-3}{\emph{Nuclear Physics B} \textbf{248} 105--122 (1984)}    

\bibitem{11}
P. de Forcrand, M. Fromm,
    \emph{ Nuclear Physics from Lattice QCD at Strong Coupling},
    \href{https://doi.org/10.1103/PhysRevLett.104.112005}{\emph{Phys. Rev. Lett.}, (2010) 104} [{\tt  	arXiv:0907.1915 [hep-lat]}]

\bibitem{12}
P. de Forcrand, J. Langelage, O. Philipsen, W. Unger,
    \emph{ The lattice QCD phase diagram in and away from the strong coupling limit},
    \href{https://doi.org/10.1103/PhysRevLett.113.152002}{\emph{Phys. Rev. Lett.} (2014) 113}[{\tt  	 	arXiv:1406.4397 [hep-lat]}]
    
\bibitem{13}
M. A. Clark, A. D. Kennedy,
\emph{Accelerating Dynamical-Fermion Computations Using the Rational Hybrid Monte Carlo Algorithm with Multiple Pseudofermion Fields},
\href{https://doi.org/10.1103/PhysRevLett.98.051601}{\emph{Phys. Rev. Lett.} (2007) 98}
[{\tt  	arXiv:hep-lat/0608015}]    
    
  
\bibitem{14}
O. Philipsen, C. Pinke, A. Sciarra, M. Bach,
\emph{CL2QCD - Lattice QCD based on OpenCL},   
\href{https://doi.org/10.22323/1.214.0038 } {\emph{PoS LATTICE2014 (2014) 038}}
[{\tt  	 	 	arXiv:1411.5219 [hep-lat]}]

\bibitem{15}
A. Sciarra, C. Pinke, M. Bach, F. Cuteri, L. Zeidlewicz, C. Schäfer, T. Breitenfelder, C. Czaban, S. Lottini, P. F. Depta. (2021),
\emph{ CL2QCD (v1.1)}
\href{https://doi.org/10.5281/zenodo.5121917}{Zenodo}


\bibitem{16}
Alessandro Sciarra,
\emph{BaHaMAS},
\href{https://doi.org/10.5281/zenodo.4577425}{\emph{Zenodo} (2021) BaHaMAS-0.4.0}
    
\bibitem{17}
P. de Forcrand, M. D'Elia,
\emph{Continuum limit and universality of the Columbia plot},
\href{https://doi.org/10.22323/1.256.0081}{\emph{PoS \textbf{LATTICE2016} (2017) 081}}
[{\tt  arXiv:1702.00330[hep-lat]}]





\end{thebibliography}
\end{document}